# B dopant evolution in Pd catalysts
# after H evolution/oxidation reaction in alkaline environment

Se-Ho Kim[a,b†,*], Su-Hyun Yoo[c,†], Leonardo Shoji Aota[a], Ayman El-Zoka[a,c], Philwoong Kang[d], Yonghyuk Lee[e], Baptiste Gault[a,c,*]

[a] Max-Planck Institut für Eisenforschung GmbH, 40237 Düsseldorf, Germany

[b] Department of Materials Science and Engineering, Korea University, Seoul 02841, Republic of Korea

[c] Department of Materials, Imperial College London, SW7 2AZ London, United Kingdom

[d] Department of Chemical and Biomolecular Engineering, Korea Advanced Institute of Science and Technology (KAIST), Daejeon 34141, Republic of Korea

[e] Fritz-Haber-Institut der Max-Planck-Gesellschaft, Berlin 14195, Germany

[†]co-first authors

[*]co-corresponding authors

**Abstract**

Introduction of interstitial dopants has opened a new pathway to optimize nanoparticle catalytic activity for, e.g., hydrogen evolution/oxidation and other reactions. Here, we discuss the stability of a property-enhancing dopant, B, introduced through controlled synthesis of an electrocatalyst Pd aerogel. We observe significant removal of B after the hydrogen evolution/oxidation reaction. Ab-initio calculations show that the high stability of sub-surface B in Pd is substantially reduced when H is ad/absorbed on the surface, favoring its departure from the host nanostructure. The destabilization of sub-surface B is more pronounced as more H occupies surface sites and empty interstitial sites. We hence demonstrate that the $H_2$ fuel/product itself favors the microstructural degradation of the electrocatalyst and an associated drop in activity.

In a recent review article, Pérez-Ramírez et al. stated that "Catalysts are not immortal", emphasizing that all catalysts degrade over time when subjected to chemical, electrical, or thermal stimuli[1], yet they are indispensable to enable the upcoming hydrogen economy. The most critical and expensive single component of the fuel cells, which can convert hydrogen gas to electricity with low cost and energy inputs[2], is the catalyst, accounting for more than 50% of the total stack cost[3]. The large-scale application of fuel cells is limited by the degradation of electrocatalysts[4,5], however, as pointed out recently, most research is focused on the catalyst's performance at all cost[6,7], and not on e.g. the deactivation mechanisms that could help design more robust and usable catalysts.

Fuel cells, including polymer electrolyte membrane[8], solid oxide[9], phosphoric acid fuel cells[10], and anion exchange membrane fuel cells (AEMFCs), use hydrogen as fuel and have attracted considerable attention[11,12]. AEMFCs operate under alkaline conditions where, on the cathode side, the oxygen reduction reaction occurs, and thus, inexpensive transition metal catalysts are used. In contrast, on the anode side, electricity is generated from hydrogen through the hydrogen oxidation reaction (HOR). At the anode, the HOR activity is more sluggish in alkaline environments compared to in acidic media[13], and there is a strong push to design more active anode catalysts[14,15].

In general, Pd exhibits a substantially low catalytic activity for HOR in alkaline media (0.05 mA $cm^{-2}_{Pd}$)[16]. However, the performance of Pd-based catalysts has been improved by decorating the surface with Ir[17] and Ru[18] clusters, alloying with Ni[19], coating on a Cu nanowire substrate[20], and doping with B[21] and $CeO_2$[22]. Among the developed variants, the Pd-$CeO_2$ catalyst, which exhibits the highest anode performance (54.5 mA $cm^{-2}_{Pd}$)[23], has emerged as one of the most promising electrocatalysts in HOR[24].

Although Pd has similar the electronic properties to Pt, the strong H- absorption and adsorption behaviours result in sluggish reaction kinetics for Pd, which limits its application in the HOR context. H binds strongly with Pd surfaces *e.g.*, $E_b^{H-Pd(111)}(\Theta_H < 0.7) = -0.69 \sim -0.34$ eV/atom, where $E_{b, H/Pd(111)}$ and $\Theta_H$ represent the binding energy of H adsorbed on Pd(111) surface and the surface coverage of H adsorbates in a monolayer (ML) unit[25–31], and forms hydrides even at low potentials. Moreover, the kinetics of the HOR in acid/base is correlated with the chemisorption energy of hydrogen on the active metallic surface; in other words, the highest activity is expected to be associated with moderate hydrogen binding energy (HBE). It has been reported that the HBE at the Pd surface decreases with increasing concentration of sub-surface B atoms in Pd[32]. Moreover, a qualitative comparison of B contents for the electrochemical absorption of H in solution shows that the H-absorption weakens at higher B-doping contents[33,34]. The B sub-surface atoms decrease the binding energy of adsorbates (HBE) on the B-doped Pd surface and strengthen the Pd structure against H attacks through Pd hydride formation, resulting in samples with more B atoms exhibiting superior kinetics in HER/HOR conditions[32].

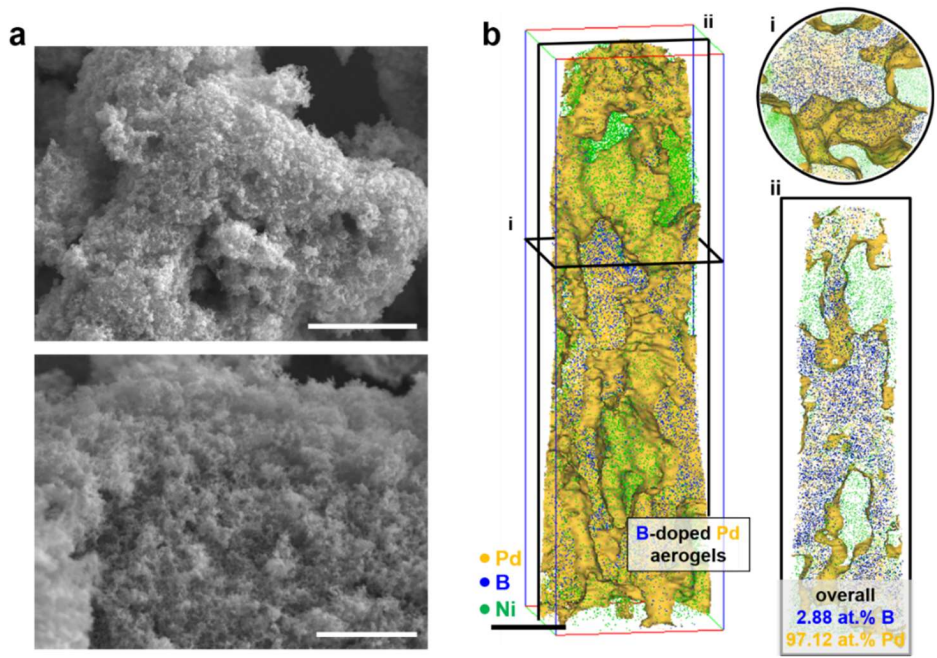

**Figure 1.** (a) SEM image of as-synthesized Pd aerogels with high (up) and low (down) magnification. Scale bars are (up) 5 and 1 μm (down). (b) 3D atom map of Pd aerogels embedded in Ni matrix. (i) and (ii) show 2-nm-thick sliced tomograms from the 3D map. The scale bar is 20 nm.

Here, we designed a model B-doped Pd catalyst to investigate the degradation mechanism during HOR. We synthesized Pd nanocatalysts using sodium borohydride ($NaBH_4$) as a reducing agent in an aqueous solution with a controlled kinetics so as to introduce over 2 at.% of B-dopants in Pd using the approach that we introduced in Ref.[32,35]. Figure 1 shows scanning electron microscopy (SEM) images and a three-dimensional (3D) reconstruction of an atom probe tomography (APT)[36,37] analysis of the as-synthesized Pd nanoparticles. The Ni matrix was used to encapsulate the porous catalyst for APT measurement following the protocol outlined in Ref.[38,39]. The Pd catalyst shows an aerogel-like network morphology, and, in the atom map, the iso-compositional surface of Pd at 25 at.% highlights the Pd catalyst surface. Consistent with our previous work[32,35], a substantial amount (2.88 at.%) of B dopants from the $NaBH_4$ reducing agent has been introduced in the Pd.

To observe the changes in the catalytic activity, first, the HOR performance associated with the as-synthesized B-doped Pd catalysts in alkaline conditions was evaluated through cyclic voltammetry (CV). The catalyst-suspension in distilled water was drop-casted onto a circular region of an Au film. Subsequently, the sample was introduced in a 150 mL solution of 0.1 M NaOH continuously purged with Ar/$H_2$ gas for 1 h. A double junction Ag/AgCl reference electrode was used with an Au coil as the counter electrode, and measurements were obtained using a Gamry potentiostat. During the Ar-saturated cycling, $H_2$ was generated in the potential below HER regime[40]. The activity in the HER/HOR region (at -0.97 V vs. Ag/AgCl) rapidly decreased for increasing cycles in the Ar environment, and gradually decreased in the $H_2$ environment, evidenced by change in onset potential owing to catalytic degradation (Figure 2a). In other words, the Pd catalyst was unstable in the $H_2$-saturated/generated HER/HOR environment in the alkaline solution. Figure 2b shows the reacted Pd catalyst after HER/HOR in an alkaline condition. The morphology of the catalyst was similar to its pristine state, and ligaments could still be observed. Noticeably, a part of the aerogel's surface was covered with a dried salt compound from the electrolyte (indicated by a white arrow in Figure 2b).

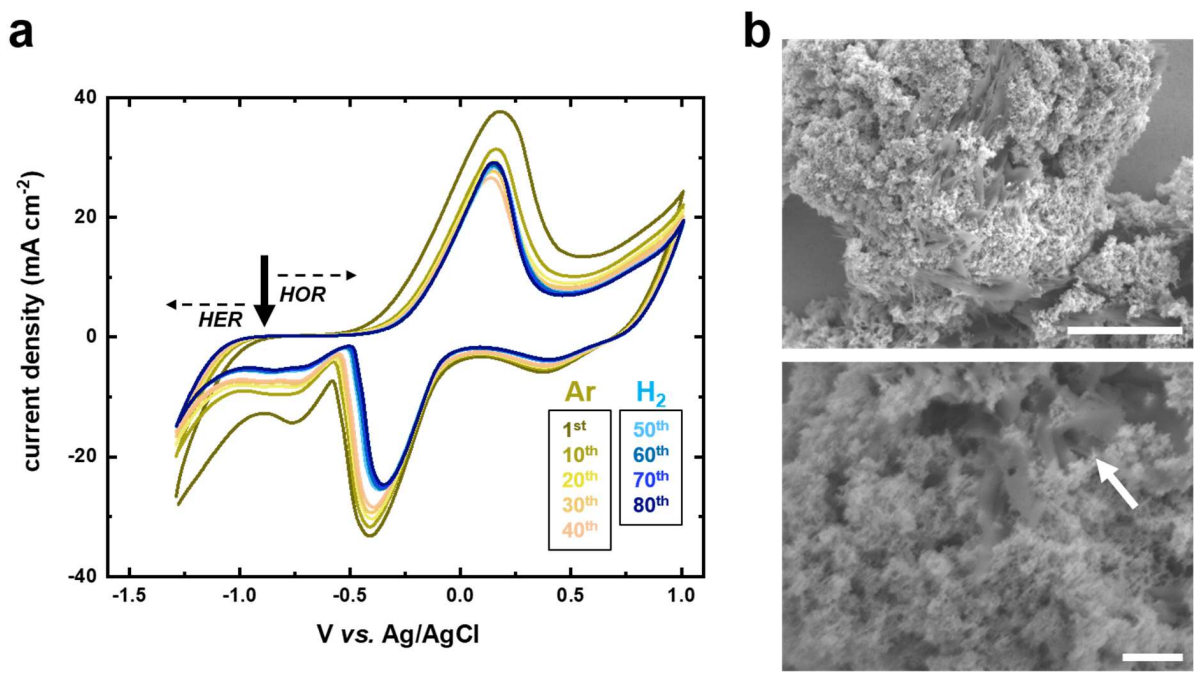

**Figure 2.** (a) CV curves at a scan rate of 50 mV·s$^{-1}$. 0.1 M of NaOH electrolyte and Au electrode were used for the HER/HOR experiment in a Ar-/$H_2$-saturated atmosphere in ambient conditions[32]. The currents were normalized according to the geometric surface area. A black arrow indicates where HER/HOR starts. (b) SEM image of post-HER/HOR Pd catalysts. The scale bars are (up) 5 and 1 μm (down).

To investigate the chemical and structural evolution of the 80-cycled catalyst, the Pd particles were collected and prepared for APT measurement, as shown in Figure 3. The atomic composition of B decreased more than twelve-fold, down to only 0.225 at.%. After prolonged electrocatalytic use, the B atoms appear to have etched away from the Pd catalyst, accompanying the decreased catalytic activity. Figure 3(i) reveals even regions in which no B atom was measured. In other datasets, the remaining B atoms appeared in the form of clusters inside the Pd. The structure, after the electrochemical test, was still in a crystalline phase as clear atomic lattices of Pd were observed from the reconstructed aerogel's grain (see inset in Figure 3(iii)). The Pearson coefficients of the frequency distribution of B ($r_B$) for the pre- and post-cycled Pd were compared to measure the degree of randomness/segregation[41], taking values of 0.21 and 0.66, respectively, indicating that B went from close to random (r=0) to close to fully segregated (r=1) upon HER/HOR cycling.

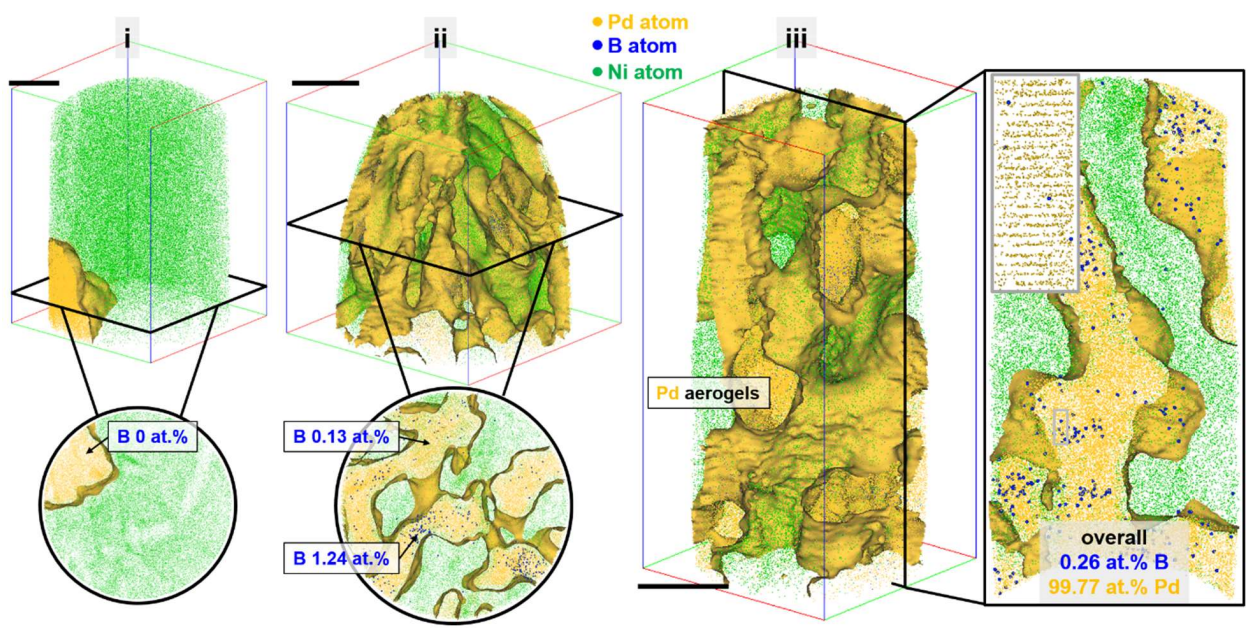

**Figure 3.** APT results of three different specimens of post-HER/HOR Pd catalysts: Pd (i) without and (ii, iii) with B. All scale bars are 20 nm.

In HOR conditions, the B-doped Pd catalyst surfaces were continuously exposed to $H_2$ gas, which results in numerous $H_2$ dissociation and adsorption events[31]. To explain the selective B dissolution, we used density-functional theory (DFT) to calculate the boron binding energies on Pd(111) surfaces as a function of H surface coverage ($\Theta_H$), as shown in Figure 4 (see Methods Section for computational details). It is noted that the surface coverage is defined as a ratio between the number of either H adsorbates or B dopants and surface Pd atoms in a ML unit, and the negatively larger binding energy indicates more exothermic adsorption. According to our previous research[32], when no H adsorbates were present (*i.e.*, $\Theta_H = 0$), the most energetically favourable sites for B dopants on Pd sub-surfaces were interstitial-octahedral sites over other high symmetric binding sites. We conducted further studies on B-doped Pd structures to understand how B binding at these sites is affected by the presence of H adsorbates.

First, as the coverage of H on the Pd surface increases, a significant decrease in the binding energy of B up to 0.84 eV/B atom is observed (as shown by the blue lines in Figure 4). When the coverage of B at octahedral sub-surface sites ($B_{octa}$) is 1 ML, the binding energy of $B_{octa}$ becomes positive at high H coverage ($\Theta_H \geq 0.75$), with a value even higher than the binding energy of B at fcc sites ($B_{fcc}$). This indicates that B dopants are energetically more favourable at surface sites than sub-surface sites when there is high H coverage on the surface with a large concentration of $B_{octa}$. In the second case, the coverage of $B_{octa}$ is fixed as 0.25 ML, which is a more representative scenario of the B dopant concentration used in the experiments. This case shows that B still binds relatively strongly in the presence of H surface adsorbates (i.e., -0.68 eV/B atom for 1 ML $H_{fcc}$ / 0.25 ML $B_{octa}$). Furthermore, considering the possibility that H atoms can easily diffuse to Pd sub-surface regions, we model additional cases where H atoms adsorb at the sub-surface octahedral sites except for the sites occupied by B atoms (see the insets of Figure 4), as shown with the purple and red lines in Figure 4. the binding energy of B is significantly reduced by up to 1.18 eV/B atom, resulting in very weak binding (*i.e.*, -0.07 eV/atom for 0.75 ML $H_{octa}$ / 0.25 ML $B_{octa}$).

The decrease in the binding energy of B indicates that i) B dopants could be largely destabilized by H surface adsorbates in regions with a higher B concentration (as indicated by the blue dotted line in Figure 4); ii) the presence of H atoms adsorbed underneath the surface can lead to very weak binding of the sub-surface B (as shown by the red and purple lines in Figure 4). These calculations suggest possibilities where B dopants would segregate on the surfaces and leave the host structure to be leached out in the solution during exposure to $H_2$ gas during HOR/HER reactions, which explains the results from APT.

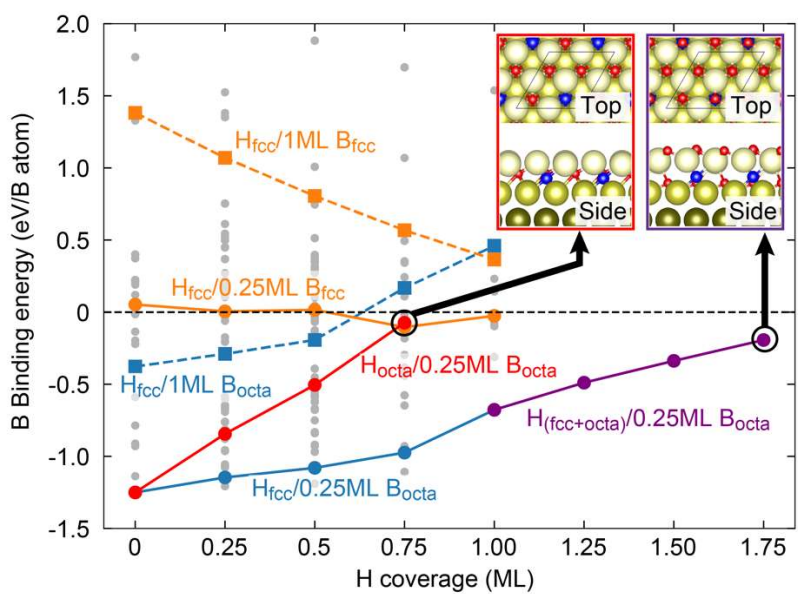

**Figure 4.** Calculated boron binding energies are plotted as a function of H surface coverage. The blue and orange lines indicate binding energies for the adsorbate-substrate structures where H are adsorbed on fcc sites ($H_{fcc}$) with B dopants at octahedral ($B_{octa}$) and fcc sites ($B_{fcc}$), respectively. The solid and dotted lines are used for $\Theta_B$ of 0.25 ML and 1 ML, respectively. The red and purple lines represent binding energies for the structures containing sub-surface H adsorbates occupied at octahedral sites ($H_{octa}$) without and with surface H adsorbates, respectively. All the grey points are the calculated binding energies of other models where B dopants occupy the other binding sites, such as top, bridge, hcp, and tetrahedral sites. The insets depict two surface models which exhibit the most weak binding of B dopants in the presence of sub-surface H adsorbates.

The presence of dopants inside electrocatalysts influences their catalytic activity, including for alkaline-HOR. Here, although B-doping was shown to increase the performance, we have observed substantial degradation of the B-doped Pd electrocatalyst during HOR, which is explained by a preferential dissolution of the B dopants in reaction conditions. Despite the relatively strong stability of sub-surface B in Pd, ab initio DFT calculations predict that both the adsorbed and absorbed H lead to a significant decrease of the B stability in Pd, which make leaving the host structure the preferred scenario. Hence, $H_2$ itself is a clear retardant of the HOR activity as it accelerates the dissolution of the property-enhancing elements added to the electrocatalysts. These insights provide hints that only increasing the catalytic activity/kinetic is not sufficient, but it must be also understood how to prevent degradation by the $H_2$ fuel to obtain truly superior HOR catalysts.

## Experimental Section

### Pd nanocatalyst synthesis

B-doped Pd catalysts were synthesized following a wet-chemical borohydride reduction method. First each potassium tetrachloropalladate (99.99%, Sigma-Aldrich) and sodium borohydride (99.99%, Sigma-Aldrich) powder was separately dissolved in distilled water. The mole ratio of $BH_4$ ions to Pd ions was set to 10. Two as-prepared solutions were mixed together, and after vigorous bubbling reaction, the Pd catalysts were synthesized. The catalysts were centrifuged to collect and were re-dispersed in distilled water for washing protocol. This process was done for three times to remove any surface remaining impurities.

### SEM analysis

After the synthesis, the catalysts were mounted on an Au electrode. Before the HER/HOR experiment, microstructural investigations were carried out using SEM (ZEISS Merlin) at an acceleration voltage of 10 kV. Then the catalyst-mounted electrode was tested for HER/HOR and, after the experiment, the electrode was again analyzed with SEM.

### APT analysis

The pre-HER/HOR Pd catalysts was encapsulated in a Ni film using co-electroplating method as described in Ref.[38]. Ga-ion sourced focused ion beam (Thermo Fischer Helios) was used to fabricate <100 nm radius APT specimens from the Pd nanoparticles/Ni composite film. APT measurements were carried out using Cameca LEAP 5000 XS system. To suppress a diffusion of the surface atoms between field evaporation sequences, the specimen temperature was set to cryogenic temperature of 60 K. A pulsed UV laser mode at a detection rate of 1%, a laser pulse energy of 50 pJ, and a pulse frequency of 125 kHz were set for data acquisition parameters.

**Electrochemical Measurements for HOR in Alkaline**

Each Pd-B powder-suspension in distilled water were drop-casted onto a fixed circular area of Au film. Loading of each sample was precisely measured by measuring the powder mass using an electric-analytical weighing system. Sample is then inserted into a 150 ml solution of 0.1 M NaOH in which Ar/$H_2$ gas was purged for 1 hr. Purging was continuous throughout the measurement. A double junction Ag/AgCl reference electrode was used, along with Au coil as a counter electrode. Measurements were carried out using a Gamry potentiostat. After the test, the B-doped Pd/Au electrodes were collected for subsequent electrodeposition and characterization studies.

**DFT calculations**

We performed DFT calculations to investigate the impact of H adsorption on the binding behavior of B dopants in Pd(111) surfaces. The Vienna Ab-initio Simulations Package (VASP) code[42,43] was used for all DFT calculations employing the projector augmented wave (PAW) method[44]. Plane-wave cutoff energy of 500 eV was used. Electronic and ionic relaxations were carried out until the total energy convergence was less than $10^{-5}$ eV. The generalized gradient approximation (GGA) due to Perdew, Burke, and Ernzerhof (PBE) was used for the exchange-correlation functional[45]. A $\Gamma$-centered (8×8×8) *k*-points grid was used for face-centered cubic Pd bulk structures, and correspondingly folded grids were used according to the size of slab structures and their supercells. The implicit solvation model implemented in the VASP code by Mathew *et al.* (*i.e.*, VASPsol), where the solvent accessible surface was taken into account from the quantum mechanically calculated charge density of solute[46], was used with setting the dielectric constant of water, $\varepsilon = 80$, to include the impact solvent has on the energetics of adsorbate-substrate models.

A supercell containing a symmetric slab of 13 Pd atomic layers (AL) with a thickness of 27.44 Å and a vacuum region of 18 Å was constructed for surface calculations. The three outermost ALs of the slab were relaxed, while the rest of ALs were fixed at their bulk positions.

To construct B-doped Pd surface structures, following our previous work where B binding behaviors were investigated[32], B dopants were positioned at surface binding sites [*i.e.*, top, bridge, face-centered cubic (fcc), hexagonal close-packed (hcp) sites] and sub-surface binding sites [*i.e.*, octahedral (octa) and two types of tetrahedral (tetra) sites] depending on its surface coverage in differently sized slab structures. The surface coverage ($\Theta$) was defined as the ratio between the number of adsorbate atoms and the surface of Pd atoms in the outmost surface layer.

Having known B-doped Pd surface structures, the impact of H adsorption on the binding behaviors of the B dopants was investigated by additionally constructing H-adsorbed B-doped Pd surface structures. First, H adsorbates were positioned at fcc surface binding sites, known as the thermodynamically most stable sites[25–31], depending on H surface coverage ($\Theta_H$), whereas H atoms are positioned atop site of B dopants for the cases where B dopants were positioned at surface binding sites. For the structures where B dopants were located at octahedral sites with $\Theta_B$ of 0.25 ML, two sets of H adsorbate-substrate models were additionally made to investigate the impact of the presence of sub-surface H adsorbates. On the one hand, H adsorbates were solely positioned at the rest of the sub-surface-octahedral sites up to 0.75 ML except for the site B dopants occupy without H surface adsorbates. On the other hand, H adsorbates were introduced at the rest of the sub-surface-octahedral sites to the structure with $H_{fcc}$ 1 ML and $B_{octa}$ 0.25 ML from 1.25 ML to 1.75 ML.

The binding energy ($E_b^B$) of B dopants as a function of H surface coverage was calculated as

$$E_b^B(\Theta_H) = \frac{1}{2N_B}\left(E_{tot}^{H/B-Pd}(\Theta_H) - E_{tot}^{H/Pd}(\Theta_H) - 2N_B\mu_B\right)$$

where $E_{tot}^{H/B-Pd}$ and $E_{tot}^{H/Pd}$ are the DFT calculated total energies of the B-doped and B-free Pd surface structures with H adsorbates, respectively. $N_B$ is the number of B dopants on one surface of the symmetric slab structure. $\mu_B$ is the chemical potential of B dopants with respect to the relevant reference phase. In this work, the reference phase for B is chosen with the rhombohedral α-phase of B which is more stable than β-phase.

## Acknowledgements

S.-H.K. and B.G. acknowledge financial support from the German Research Foundation (DFG) through DIP Project No. 450800666. S.-H.Y. acknowledge funding from the European Union's Horizon 2020 research and innovation programme under the Marie Sklodowska - Curie grant agreement No. 101034297.